\documentclass{osa-article}\usepackage{graphicx,color}
\journal{oe}
\usepackage{bm}
\usepackage{float}
\usepackage{ulem}
\usepackage{amsmath}
\usepackage{mathtools}
\usepackage[utf8]{inputenc}
\usepackage{jabbrv}

\begin{document}

\title{Optimal conditions for multiplexing information into ring-core optical fibers}

\author{S.~Rojas-Rojas,\authormark{1,2} G.~Ca\~{n}as,\authormark{3} G.~Saavedra,\authormark{4,*} E.~S.~G\'{o}mez,\authormark{1,2} S.~P.~Walborn,\authormark{1,2} G.~Lima,\authormark{1,2} }

\address{\authormark{1}Departamento de F\'isica, Universidad de Concepci\'on, 160-C Concepci\'on, Chile\\
\authormark{2} {Millennium Institute for Research in Optics, Universidad de Concepci\'on, 160-C Concepci\'on, Chile} \\
\authormark{3}Departamento de F\'isica, Universidad del B\'io-B\'io, Collao 1202, 5-C Concepci\'on, Chile\\
\authormark{4}Departamento de Ingenier\'ia El\'ectrica , Universidad de Concepci\'on, 160-C Concepci\'on, Chile}

\email{\authormark{*}gasaavedra@udec.cl} 

\begin{abstract}
In optical communications, space-division multiplexing is a promising strategy to augment the fiber network capacity. It relies on modern fiber designs that support the propagation of multiple spatial modes. One of these fibers, the ring-core fiber (RCF), is able to propagate modes that carry orbital angular momentum (OAM), and has been shown to enhance not only classical, but also quantum communication systems. Typically, the RCF spatial modes are used as orthogonal transmission channels for data streams that are coupled into the fiber using different Laguerre-Gaussian (LG) beams. Here, we study the optimal conditions to multiplex information into ring-core fibers in this scheme. We determine which are the most relevant LG  beams to be considered, and how their coupling efficiency can be maximized by properly adjusting the beam width with respect to the fiber parameters. Our results show that the coupling efficiency depends upon the OAM value, and that this can limit the achievable transmission rates. In this regard, we show that LG beams are not the optimal choice to couple information into RCF. Rather, another class of OAM-carrying beam, the perfect vortex beam, allows for nearly perfect coupling efficiencies for all spatial modes supported by these fibers.
\end{abstract}

\section{Introduction}

Over the last 50 years, the capability of fiber optic technology to deal with an ever growing demand for higher communication rates has fueled a revolution in telecommunication and networking industries, as well as in science and engineering. Using available techniques for time-, polarization-, and wavelength-division signal multiplexing, high capacity communications systems have been implemented resorting to single-mode fibers (SMFs)\cite{Zhu, Kan,Gnauck, Galdino, Ionescu:20, Hamaoka, Renaudier}. However, nowadays the information capacity carried by SMFs is rapidly approaching its physical limit \cite{Winzer, Richardson_1, Essiambre}, a problem known as the ``capacity crunch''. One of the main proposals to overcome this limiting issue is the use of multiple spatial channels to multiplex signals into optical fibers, in addition to the aforementioned techniques \cite{Richard13,Winzer}. To achieve this, novel optical fibers that support propagation of several spatial modes have been under development. To date, at least three main fiber types can be found, namely, few mode fibers (FMFs) \cite{Sillard, Rademacher}, multi-core fibers (MCFs) \cite{Saitoh}, and ring-core fibers (RCFs) \cite{RCF}.

\begin{figure}[t]
 \centering
 \includegraphics[width=.95\textwidth]{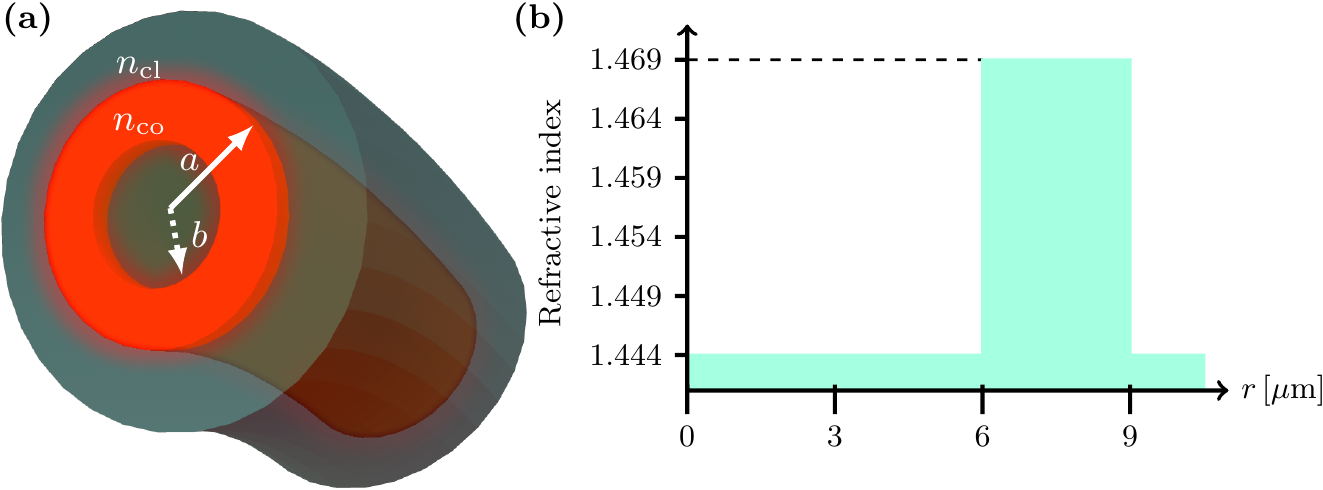}
  \caption{(a) Schematics of a ring-core fiber. Light propagates through the annular core with internal and external radius $b$ and $a$, respectively, delimited by refractive index $n_{\rm co}$ embedded on a cladding with refractive index $n_{\rm cl}$. (b) Radial profile of the refractive index for the fiber parameters used in our study.}\label{fg:intro}
\end{figure}

Ring-core fibers (see Fig.~\ref{fg:intro}) have been successful in this context, with demonstrations showing that the spatial modes of these fibers can be used to enhance not only classical \cite{Bozinovic, Nejad, RCF_Wang}, but also quantum communication links \cite{Bacco, Boyd, Cao:20,Glima20B}. The spatial modes of a RCF can carry orbital-angular-momentum (OAM), and are usually excited using independent data streams propagating in different free-space Laguerre-Gaussian (LG) beams \cite{Bozinovic,Zhang:20,Zhu:18}. 
However, since there are several spatial modes supported by the fiber, the optical configuration adopted to couple the beams into the RCF may lead to drastic differences in the coupling efficiencies of each input signal. For instance, a LG beam - characterized by radial (p) and azimuthal ($\ell$) numbers -  is a helically phased beam composed of an azimuthal phase term $e^{i\ell \phi}$, where each photon carries OAM $ \ell\hbar$, with $\ell$ being the topological charge of the beam and $\phi$ its azimuthal angle \cite{Allen,Padgett}. The LG beam width depends on both of these numbers. For constant radial number, the ring radius size scale with $\sqrt{|\ell|}$. The spatial eigenmodes of the RCF, on the other hand, have a fixed width parameter that is defined only by the core radius, and consequently an asymmetry on the coupling efficiencies as a function of $\ell$ is observed for a given optical configuration. This results in space-division multiplexing communication schemes with channels that may have drastically different overall transmissions, which in general decreases the achievable communication rates. For quantum communication, asymmetrical coupling efficiencies result in undesired state transformations performed by the fiber, resulting in high quantum error rates. 

To overcome such limitations, we study the optimal conditions to multiplex information into a RCF. First, we show which are the most relevant LG beams to be considered, and how their coupling efficiency can be maximized by properly adjusting the beam width with respect to the fiber core radius. Then, we show that LG beams are not optimal to couple signals multiplexed in the spatial domain into a RCF. As an alternative to the usual LG  beams, we consider the perfect vortex (PV) beams introduced by Ostrovsky et al. \cite{Ostrovsky_13,Ostrovsky_14}, and we show that these PV beams allow for nearly perfect coupling efficiency for all spatial modes supported by RCFs.


\section{The spatial modes of ring-core optical fibers}

\begin{figure}[t]
\centering
 \includegraphics[width=.97\textwidth]{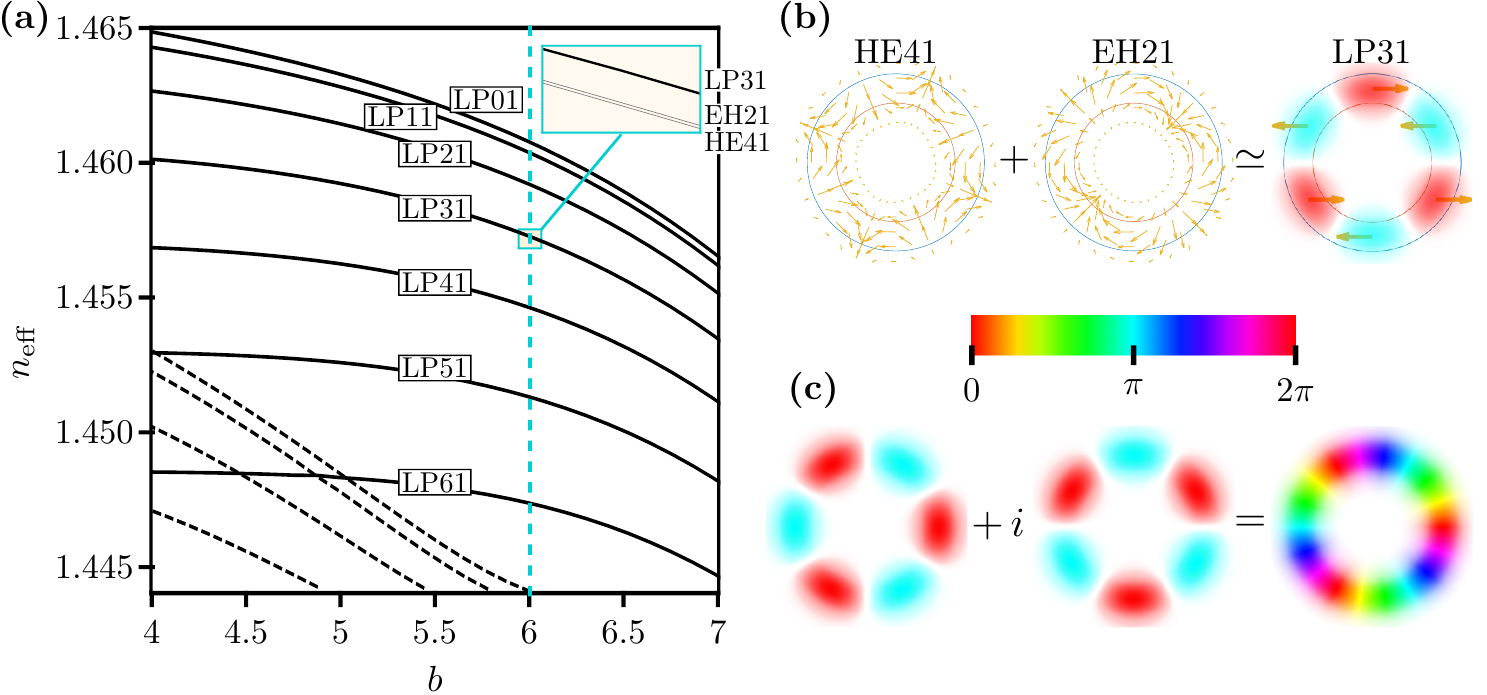}
 \caption{(a) Effective refractive index of the linearly polarized modes supported by the RCF, as a function of the internal radius $b$. The vertical blue line marks the particular configuration $b=6\,\mu$m used in our calculations, which supports the seven modes indicated by the labels, with the same radial order $m=1$ (higher-order modes are represented by the dashed curves). The example in the inset shows an enlarged region of the plot including the exact (vector) modes. (b) Example of a combination of exact vector modes giving rise to a linearly polarized mode in the limit $n_{\rm co}\simeq n_{\rm cl}$. (c) A complex superposition of orthogonal LP modes allows to encode OAM. In this example the topological charge is $\ell=3$.}\label{fg:modos}
\end{figure}

To study the coupling efficiency between a RCF and different spatial modes propagating in free-space, first we need to determine the bound modes of the fiber. The system under consideration is a ring core fiber, illustrated in Fig. \ref{fg:intro}~(a).  Let $z$ be the direction corresponding to the longitudinal axis of the RCF. The electric and magnetic components of the $j$-th bound mode carried by the fiber can then be expressed as ${\bf e}_je^{i\beta z}$ and ${\bf h}_je^{i\beta z}$ respectively, where amplitudes ${\bf e}_j$ and ${\bf h}_j$ solve the vector eigenvalue equations derived from the source-free Maxwell equations~\cite{Snyder}. The corresponding eigenvalue $\beta_j$ is the propagation constant of the mode. Depending on the symmetry of the particular problem, exact solutions of the vector equations can be found~\cite{vector}.  When the contrast $\Delta n$ between the core ($n_{\rm co}$) and cladding ($n_{\rm cl}$) refractive indices is low enough, different subsets of vector (exact) modes become nearly degenerate. This regime, commonly referred to as the {\it weakly guiding approximation} \cite{SnyYoung}, is attainable with standard fabrication techniques and enables linear combinations of the exact solutions to become bound modes of the fiber. In particular, the fiber sustains {\it linearly polarized} (LP) modes whose longitudinal field components are small compared to the transverse components, which keep the same direction of polarization across the transverse section (unlike the exact vector modes). This last property is related to the LP modes being solutions of the scalar equation $\nabla_t^2\Psi_\ell+(k^2n^2-\beta_\ell^2)\Psi_\ell=0$, derived from the exact vector equations in the limit $n_{\rm co}\simeq n_{\rm cl}$. Therefore, in the weakly guiding approximation the LP modes form a basis, where for each propagation constant $\beta_\ell$ other than the fundamental one, two orthogonal states of different parity (i.e. different variation with the azymuthal angle $\phi$). Indeed, linear combinations of such degenerate basis states are also eigenmodes of the fiber with the same eigenvalue, which makes it possible to encode OAM using LP modes with azimuthal dependence $\exp(i\ell\phi)$, depicted in Fig.~\ref{fg:modos} (c). 
If the external cladding is taken to have a finite extension, their spatial profile can be expressed as
\begin{equation}\label{eq:lp}
\Psi_\ell(r,\phi)=S(\ell\phi)
 \begin{cases}
         C_1{\rm I}_\ell(wr) & 0\leq r < b\,,\\
    A_1{\rm J}_\ell(ur)+A_2{\rm Y}_\ell(ur) & b\leq r < a\,,\\
    C_2{\rm K}_\ell(wr)+C_3{\rm I}_\ell(wr) & a\leq r \leq c\,,
 \end{cases}
\end{equation}
where $S(\ell\phi)$ is either $\cos(\ell\phi)$ or $\sin(\ell\phi)$ depending on the mode parity, while $J_\ell$ $Y_\ell$, $K_\ell$ and $I_\ell$ are the Bessel functions of the first and second kind. Real coefficients $A_i$ and $C_i$ are determined by the condition that fields described by Eq. \eqref{eq:lp} must be continuous and smooth across all the fiber section, and null in the cladding border $r=c$, in order to solve the scalar wave equation. These requirements results in the following characteristic equation for $\beta_\ell$:
\begin{equation}
\begin{split}
 &\frac{{\rm I}'_\ell(wb){\rm J}_\ell(ub)-\frac{u}{w}{\rm J}'_\ell(ub){\rm I}_\ell(wb)}{{\rm I}'_\ell(wb){\rm Y}_\ell(ub)-\frac{u}{w}{\rm Y}'_\ell(ub){\rm I}_\ell(wb)}\\
 &=\frac{\left({\rm K}'_\ell(wa)-\frac{{\rm K}_\ell(wc)}{{\rm I}_\ell(wc)}{\rm I}'(wa)\right){\rm J}_\ell(ua)-\frac{u}{w}{\rm J}'_\ell(ua)\left({\rm K}_\ell(wa)-\frac{{\rm K}_\ell(wc)}{{\rm I}_\ell(wc)}{\rm I}(wa)\right)}{\left({\rm K}'_\ell(wa)-\frac{{\rm K}_\ell(wc)}{{\rm I}_\ell(wc)}{\rm I}'(wa)\right){\rm Y}_\ell(ua)-\frac{u}{w}{\rm Y}'_\ell(ua)\left({\rm K}_\ell(wa)-\frac{{\rm K}_\ell(wc)}{{\rm I}_\ell(wc)}{\rm I}(wa)\right)}\,,
\end{split}
\end{equation}
with $w^2=\beta_\ell^2-k_0^2n_{\rm cl}^2$ and $u^2=k_0^2n_{\rm cl}^2-\beta_\ell^2$ being the fiber parameters. Note that in the limit $c\rightarrow \infty$ we recover the characteristic equation for LP modes in annular cores \cite{lps,lps2}. We start by computing the bound modes of a RCF with external radius $a=9.0\,\mu$m for different internal radii (cf.~\cite{Kasahara}). The fiber material is taken to be fused silica, such that $n_{\rm cl}=1.444$ when the wavelength of the incident light is $\lambda=1\,550$\,nm. The refractive index contrast is $\Delta n=0.025$. We first solve the scalar equation to obtain the LP modes of the fiber. Our results for the effective refractive index $n_{\rm eff}$ (proportional to the propagation constant) of the modes are shown in Fig.~\ref{fg:modos} (a). In order to ensure that the LP modes can be used to encode OAM, we need to confirm the validity of the weakly guiding approximation for the parameters in our analysis. To evaluate this, we solved the exact problem and obtained the $n_{\rm eff}$ curves for the vector solutions. The exact results are very well approximated by the LP modes, so the curves overlap in the full picture. Furthermore, we estimate the time spread of the LP modes to be $\sim0.1\,$ns per kilometer. For an internal radius $b$ equal to $6.0\,\mu$m, (vertical blue line in the figure) we find that the fiber supports thirteen modes LP$_{\ell 1}$: the fundamental $\ell=0$ mode and two parities for each $\ell$ between 1 and 6. The second index $m=1$ indicates that all the modes are first-order regarding the radial distribution of the field amplitude.  In Ref. \cite{RCF_Wang}, a RCF with an internal radius of $6.0\,\mu$m was used for space-division multiplexing, and analysis in the following sections are made with this configuration. The example in Fig. \ref{fg:modos} (b) illustrates how LP modes arise as linear combinations of the vector modes in the limit where these become degenerate. In Fig. \ref{fg:modos} (c) we show a combination of LP modes allowing to encode OAM. For each OAM order, the topological charge is given by $\pm |\ell|$ ($\ell=3$ in the example) so it can be excited by a free-space beam with the same $\ell$ and linear polarization. Note that this is not possible when the refractive-index contrast is high, such as in fibers with doped cores \cite{Brunet:14}. In that regime, OAM modes must be constructed as phase-shifted combinations of even and odd vector modes, so the propagated field has circular polarization. Finally, we note that for certain quantum information protocols the relative time delay between modes, induced by the difference in their effective refractive index, may be relevant. In our case, the delay between the fundamental and the highest-order mode is 20.97\,ps (10.82\,ns) after a transmission distance of 50\,cm (1\,km). We leave detailed study of the temporal delay of these modes for a future study, and remark that the relative delays can be corrected at the receiver or transmitter, if necessary.

\section{Coupling efficiency between the fiber and free-space spatial modes}
The OAM modes in a RCF can be used as orthogonal carriers to multiplex data streams. Typically, free-space modes are used to excite OAM fiber modes. We shall now study the coupling conditions to optimally multiplex information into ring-core fibers, using LG modes and PV beams.   

\subsection{Using LG beams}

\begin{figure*}[t]
 \centering
 \includegraphics[width=0.95 \textwidth]{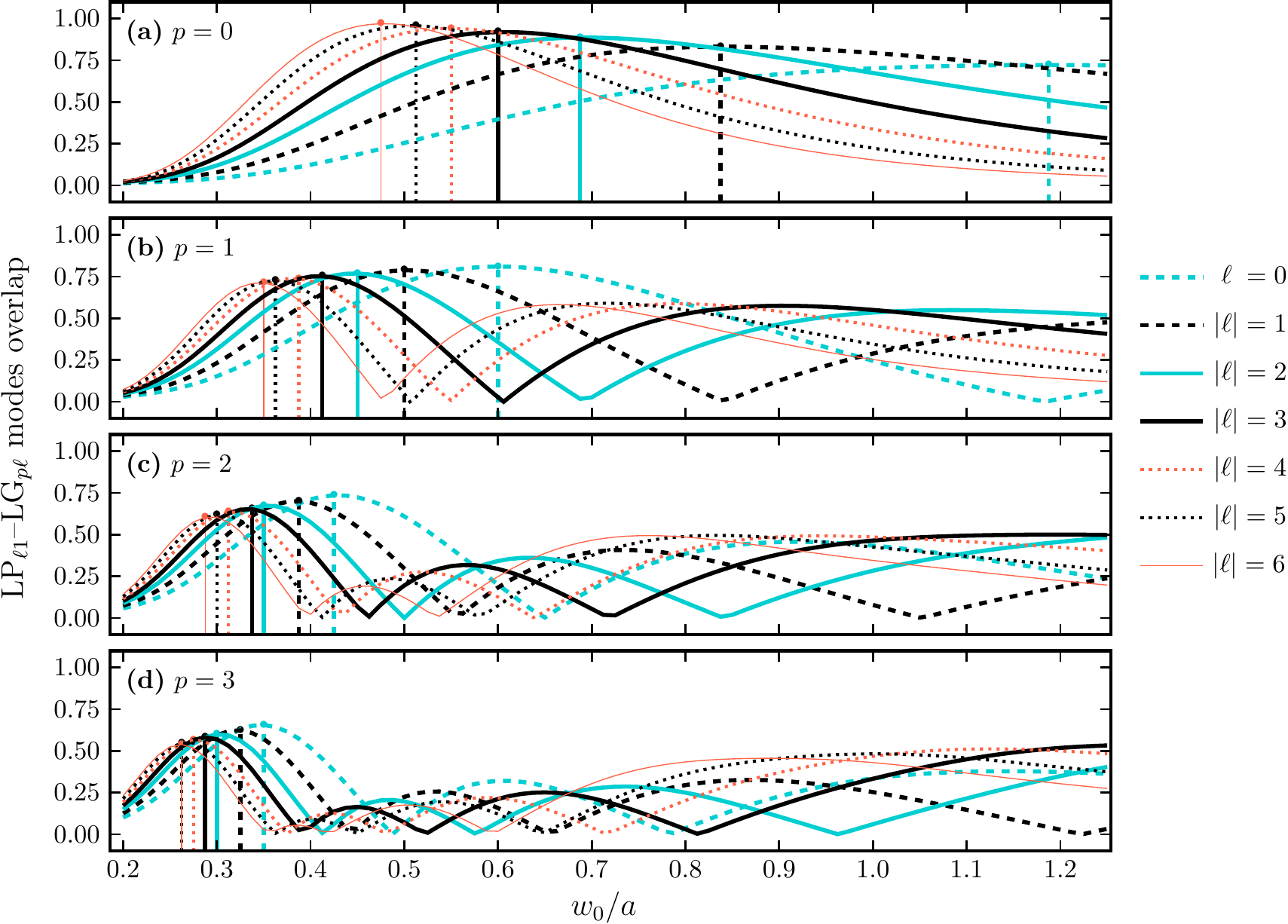}
 \caption{Overlap between the LP$_{\ell 1}$ modes sustained by the fiber and the matching Laguerre-Gaussian modes LG$_{p\ell}$ up to $p=3$, as a function of the ratio between the width $w_0$ of the LG-modes and the external radius $a$ of the fiber core.}\label{fg:overlaps}
\end{figure*}

LG beams are a suitable choice for the adopted free-space modes to excite the RCF modes since they share the cylindrical symmetry of the fiber. Moreover, they are eigenmodes of first-order optical systems with cylindrical symmetry, such as free space propagation or spherical lenses. These modes are solutions of the paraxial Helmholtz equation, and are given by~\cite{Allen_Padgett}:
\begin{equation}\label{eq:lg}
  \text{LG}_{p\ell}=M_{p\ell}\left(\frac{2r^2}{w_0^2}\right)^{\frac{|\ell|}{2}} L_p^{|\ell|}\left(\frac{2r^2}{w_0^2}\right)\exp\left(-\frac{r^2}{w_0^2}\right)\,\exp(i\ell\phi),
\end{equation} 
where, $L_p^{|\ell|}$ are the associated Laguerre polynomials, $w_0$ defines the width of the beam and $M_{p\ell}$ is a normalization factor.

Using LG beams to couple OAM modes into a RCF has been successfully demonstrated, and was used in~\cite{bozinovic2013terabit,RCF_Wang}. Despite this, coupling multiple OAM with different topological charge remains a practical challenge, as the diameter of the LG mode is proportional to $\sqrt{|\ell|}$. In practice, this means that one cannot simultaneously optimize the coupling of different free space modes into the fiber using the same optical configuration. In~\cite{bozinovic2013terabit}, optical modes use the same topological charge $\ell=\pm 1$ with opposite wavefront rotation directions, and in~\cite{RCF_Wang} OAM modes where multiplexed with contiguous $\ell=\{+4,+5\}$ to simplify the coupling setup. Note that both references use systems with 2 spatial modes or dimensions.

To analyze the coupling efficiency of LG beams in a RCF we use the projection of the modes onto the LP basis and viceversa. As described in the previous section, OAM modes within the RCF can be generated as a linear combination of LP modes. We consider the following figure of merit, which measures the overlap between the modes:
\begin{equation}\label{eq:overlap}
 \eta = \left| \iint {\rm LG}_{p\ell}(r,\phi) \Psi_{\ell}^\ast(r,\phi)\, dA\,\right |.
\end{equation}

It follows directly that LG and LP modes can be matched only if they have the same azimuthal index $\ell$, in accordance with the definition of $\eta$, since they both have the same angular dependence $\exp(i\ell\phi)$. 
The relationship in Eq.~\eqref{eq:overlap} was used by Br\"uning \textit{et al.} in Ref.~\cite{overlap} to study the overlap of the LG modes with the eigenmodes of a step-index fiber. Unlike that case, modes of the RCF can be matched to LG modes having different radial order due to the ring shape of the core. Therefore, for each LP mode we evaluate the overlap with the first four radial orders ($p=0,1,2,3$) of the corresponding LG modes. The overlap between LG and LP modes can be characterized by a single parameter--the ratio $w_0/a$ between the beam width $w_0$ of the LG mode and the external core radius $a$ of the RCF, and is shown in Fig.~\ref{fg:overlaps}. 
Vertical lines show the maximum coupling efficiency for a given order $\ell$. Each sub-figure presents a different radial order $p$ of the LG beams. As the azimuthal order of the LG mode is increased, the ratio $w_0/a$ needs to be decreased in order to couple light into the fiber with maximum efficiency. In general, higher coupling efficiencies are observed using LG beams with radial order $p=0$. However, large differences are observed in the optimal $w_0/a$ ratio for different values of $\ell$. For example, for $p=0$, a $w_0/a$ of $1.2$ is required to optimally couple $\ell = 0$ into the RCF, while $w_0/a$ of $0.47$ is needed for $\ell = 6$. Alternatively, using $p=3$ results in lower coupling efficiencies, however the optimal $w_0/a$ ratios of $0.35$ and $0.25$ for $\ell = 0$ and $6$ are much closer. For radial orders $p>0$, multiple peaks in the coupling efficiency are observed as $w_0/a$ is increased. This is due to the fact that LG modes can have multiple rings, and as the beam width is increased the inner rings are coupled into the RCF. However, the highest overlap is observed for the outer ring of any LG mode, which usually has the highest intensity. 

The maximal coupling efficiency that can be achieved for each optical configuration studied is highlighted in Table \ref{tab:max_efficiency}. Note that for the $p = 0$ the efficiency increases together with the azimuthal number, while for $p>0$ the opposite effect is observed. In the former case, this is due to the fact that the rings get narrower as a function of $|\ell|$. In the latter case, as discussed previously, in the optimal coupling scenario only the external ring of the LG mode is coupled into the fiber for $p>0$, leading to coupling losses.

\begin{table}[!b]
\begin{center}
\begin{tabular}{c|cccc|}

  $\eta_{\rm max}$ & LG$_{0\ell}$ & LG$_{1\ell}$ & LG$_{2\ell}$ & LG$_{3\ell}$ \\ \hline

  LP$_{01}$ & 0.7214 & 0.8101 & 0.7353 & 0.6553  \\

  LP$_{11}$ & 0.8314 & 0.7878 & 0.7000 & 0.6235  \\

  LP$_{21}$ & 0.8860 & 0.7686 & 0.6729 & 0.6000  \\

  LP$_{31}$ & 0.9188 & 0.7532 & 0.6525 & 0.5818  \\

  LP$_{41}$ & 0.9411 & 0.7372 & 0.6355 & 0.5657  \\

  LP$_{51}$ & 0.9568 & 0.7258 & 0.6205 & 0.5484  \\

  LP$_{61}$ & 0.9690 & 0.7199 & 0.6039 & 0.5389  \\

\end{tabular}
\end{center}
\caption{Maximum overlap between different pairs of $LP_{\ell 1}$ and $LG$ modes.}
\label{tab:max_efficiency}
\end{table}

\subsection{Using PV beams}

\begin{figure}[t]
 \centering
 \includegraphics[width=0.95 \textwidth]{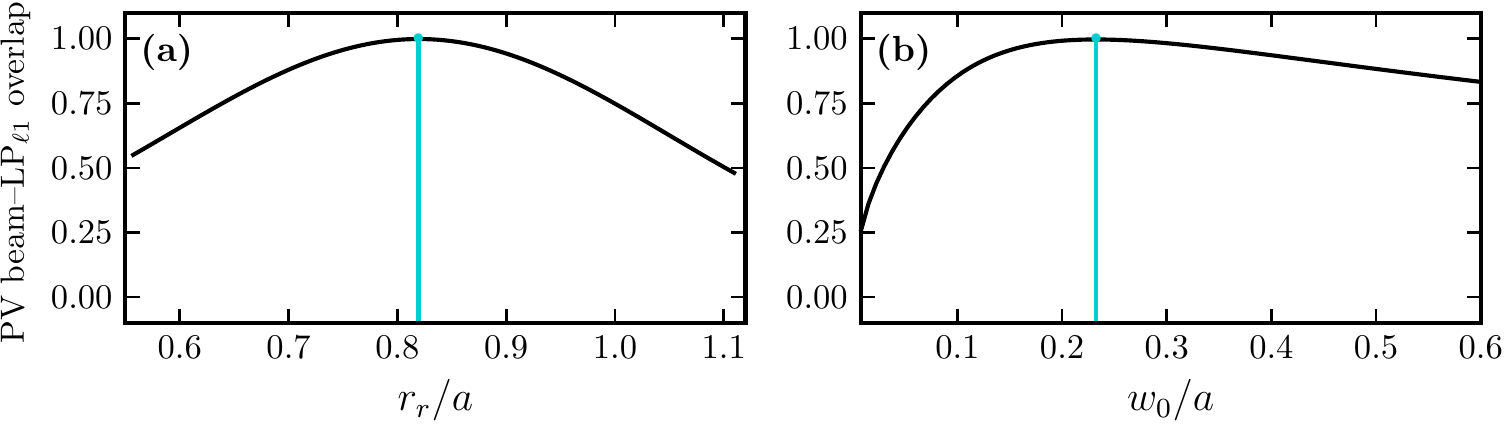}
 \caption{Overlap between the LP modes of the fiber and PV beams as a function of its radius $r_r$ (a) and its width $w_0$ (b), in units of $a$. Vertical blue lines indicate the parameters for optimal coupling. The coupling is independent of the topological charge.}\label{fg:pvb}
\end{figure}

Our results show that the coupling efficiency between LG and LP modes strongly depends on the topological charge $\ell$, even for the case of constant radial index $p$. This must be considered when coupling multiple OAM modes into ring core fiber, and leads to a trade-off between optimality and homogeneity in terms of the coupling efficiencies. To eliminate this $\ell$-dependence, we now consider PV beams, which have a field profile that is more convenient for this type of application. The PV beams are obtained as Fourier transformations of Bessel-Gaussian beams, and have a transverse field distribution given by \cite{pvb}:
 
\begin{equation}\label{eq:pvb}
  PV_{\ell}\simeq i^{\ell-1} \frac{w_g}{w_0}\exp(i\ell\phi)\exp\left(-\frac{(r-r_r)^2}{w_0^2}\right),
\end{equation}
where $w_g$ is the beam width of the Gaussian beam which is used to confine the Bessel beam, $w_0$ is the beam width at the focus plane ($w_0=2f/kw_g$), and the annular profile of PV beams have thickness and radius of the ring equal to $2w_0$ and $r_r$, respectively. Thus, as long as $r_r$ is large enough for the approximation of Eq.~(\ref{eq:pvb}) to be valid, it is possible to set the field amplitude to have the desired transverse ``ring" profile that is independent of the value $\ell$. We note that PV beams have also been used to demonstrate the propagation of OAM modes through specially designed ring core and air core fibers, which support up to $36$ and $10$ OAM modes, respectively \cite{P_Vaity, Brunet:14}. In our case, the linear polarization and the sign of the topological charge allows the fiber to support 26 modes.  Different to the case of conventional few- and multi-mode fibers, in the RCF the spatial modes are confined within the annular core in such a way that the radial profile of the LP modes only varies slightly with $\ell$, but is determined by the radii parameters $a$ and $b$. Thus, it is possible to find a single $w_o$ and $r_r$ which optimally couples each LP$_{\ell1}$ to a PV beam PV$_{\ell}$ with the same topological charge. This is shown in Fig.~\ref{fg:pvb}, where we show the overlap between the PV and the LP modes obtained by using the PV beam of Eq. \eqref{eq:pvb} in relation \eqref{eq:overlap}. In this case, an average coupling efficiency of $0.9959$ is achieved for the ratios  $r_r/a=0.83$ and $w_0/a=0.235$.  If instead we look for specific PV beam optimized for each $\ell$, the maximum overlap is 0.9986 for $\ell=3$, close to the 0.9989 benchmark attainable with exact vector modes. This coupling efficiency outperforms all cases considered for the LG beams. For instance, the best coupling efficiency of LG beams is achieved for $p=0$, which is about $5\%$ lower than the coupling efficiency of PV into RCF modes. Furthermore, for $p>0$ the coupling efficiency is $25\%$ to $50\%$ lower than PV beams (see Fig.~\ref{fg:overlaps}).

\subsection{LG vs PV beams coupling efficiencies}

\begin{figure}[ht]
 \centering
 \includegraphics[width=.94\textwidth]{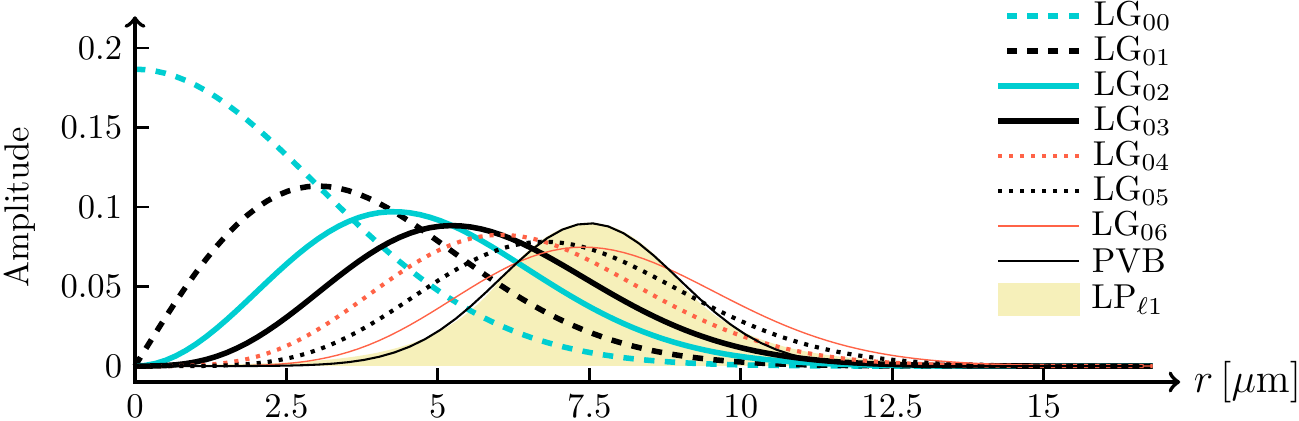}
 \caption{(a) Variation in the radial profile of the LG$_ {0\ell}$ modes as the topological charge $\ell$ is increased, for a fixed beam width $w_0=0.475$. Their radius scales as $\sqrt{\ell}$, so the maximal overlap with the LP modes (orange area)  is achieved with $\ell=6$ for the chosen $w_0$. 
 }\label{fg:pvbvsLG}
\end{figure}

To further compare the use of both LG and PV beams to excite OAM modes in a RCF, we now use the radial profile of the modes and examine how they compare to those of the bound modes of the fiber. Figure~\ref{fg:pvbvsLG} (a) shows the radial amplitude profiles of a variety of studied beams. As discussed above, the annular structure strongly determines the radial profile of the LP modes (the average is shown by the beige region in the figure) so a single PV beam can be found (black curve) which optimally couples to all LP modes, with an average overlap of 0.9959. Profiles of the LG beams are shown in the figure for $w_0/a=0.475$, which corresponds to the highest coupling efficiency achieved in Fig. \ref{fg:overlaps} (for $\ell=6$). Visual inspection of the overlap between amplitudes clearly shows that sub-optimal coupling is achieved for every LG mode when the beam width is fixed. LG modes LG$_{06}$ and LG$_{05}$ have similar amplitude profiles, and thus similar efficiency is observed (see Fig.~\ref{fg:modos} (a)). However, LG modes with lower azimuthal order deviate greatly from the LP mode profile.
On the other hand, since the radial profile of the PV beam is independent of the azimuthal charge, the same overlap is observed between PV$_\ell$ and the bound mode LP$_{\ell 1}$ supported in the fiber. Despite this, we can observe that the overlap between the PV and the LP modes is not perfect. The radial profile of the PV beam deviates slightly from the LP mode because the former is explicitly defined to have a Gaussian profile around $r_r$. Since the LP modes must satisfy the boundary conditions for the parallel components of the electromagnetic field, its radial profile exhibits a different decay in the inner and the outer cladding, described by different kinds of Bessel functions in each region. 

\subsection{Achievable dimension of quantum communication channels}
To use OAM as a viable candidate to solve the "capacity crunch" in optical fiber communication systems, and to expand quantum communication systems, a large number of OAM modes need to be multiplexed into a single RCF. As studied here, the use of LG beams to multiplex OAM into a RCF will lead to different transmission losses for each spatial channel. For classical communications links this will result in different quality of transmission for each channel, limiting the amount of information that can be encoded into it.
\par
In quantum information, it is well known that some protocols can be more robust when using quantum states in higher dimensions (qudits) \cite{QC_1,QC_2,QC_3,QC_4}. In the current scenario, a typical approach is to encode a $d$-dimensional qudit state as a single photon in a superposition of LG modes with different OAM $\ell$ and fixed radial order $p$ \cite{OAM_1,OAM_2,OAM_3}. Here it is necessary to couple not only the individual OAM basis states into the RCF, but also superposition states with reasonable fidelity. Thus, one must search for a ratio parameter $w_0/a$ that gives the same coupling efficiency for several OAM modes. For instance, a mode overlap of $\sim 0.95$ can be achieved for the LG modes with $p=0$ and $\ell=\pm 4, \pm 5, \pm 6$ for the ratio $w_0/a=0.52$, allowing for a fairly high and homogeneous coupling efficiency for the basis elements of a 6-dimensional quantum state  (see Fig.~\ref{fg:overlaps}a). Similar situations for six-dimensional states occur for radial orders $p>0$, but with a coupling efficiency less than $75\%$ for a ratio parameter $w_0/a$ between $0.25$ and $0.35$ (see Fig.~\ref{fg:overlaps}). Consequently, the dimension of the quantum state encoded in the LG modes is restricted by the coupling efficiency between LG and RCF modes, instead of the number of allowed propagation modes allowed by the RCF. We note that the overall effect of coupling into the RCF is a non-unitary filtering operation on the qudit state.  This operation could of course be corrected at the expense of further losses. Therefore, though we identify 13 different eigenmodes in the RCF we study, a 13-dimensional quantum states cannot be transmitted into the RCF without a drastic loss in fidelity, efficiency or both. 
\par 
On the other hand, the PV beams can be used to encode quantum states in higher dimensions, and to encode classical channels in SDM systems. Due to the characteristic property of the PV beams, which is that the beam shape is independent of the topological charge $\ell$, PV beams present a constant coupling efficiency for all OAM modes supported by a RCF. In this case, the dimension of a quantum state and the number of spatial channels are limited ultimately by the propagation modes allowed by the RCF instead of the coupling efficiency of the free-space beams into the fiber. The use of PV beams will lead to improved transmission systems, and simplify the optical setup to generate high order quantum and classical communication systems using RCF.

\section{Conclusions}
The use of orbital angular momentum to generate spatially multiplexed channels in optical fiber has the potential to reduce the impact of the capacity crunch in classical communications fibers, and to increase the efficiency of quantum communication links. By evaluating the overlap between free space optical beams capable of carrying OAM and the spatial modes supported by a ring-core fiber we have computed the coupling efficiency between them, for different beam parameters. We show that for Laguerre-Gaussian input beams, the coupling efficiency depends not only upon the beam width and fiber core diameters, but also upon the OAM value of the beam.  This leads to a decrease in communication capacity, as some OAM channels will be coupled worse than others. Presumably, one could resort to  much more complex optical systems to achieve homogeneous coupling efficiencies. As an alternative solution, we investigate the use of perfect vortex beams as input to the RCF. We show that in this case the coupling efficiencies are nearly independent of the OAM value, rendering these beams as much more suitable for multiplexing OAM channels from free space into a ring-core fiber.  We expect these results to play an important role in space-division multiplexing of both classical and quantum optical information.

\section*{Acknowledgments}
This work was supported by Fondo Nacional de Desarrollo Científico y Tecnológico (Fondecyt 1190933, Fondecyt 1190710, Fondecyt 1190901, Fondecyt 1200266, and Fondecyt 1200859), and by ANID - Millenium Science Initiative Program - ICN17\_012. S.R.R. acknowledges support from Fondecyt 3180752. 
\section*{Disclosures}
The authors declare no conflicts of interest.
\bibliography{rcf}

\end{document}